\documentclass[prl,aps,showpacs,twocolumn]{revtex4}%
\usepackage{graphicx}
\usepackage{amsmath}
\usepackage{amsfonts}
\usepackage{amssymb}%
\providecommand{\U}[1]{\protect\rule{.1in}{.1in}}
\begin{document}
\title{A Solvable Model for Polymorphic Dynamics of Biofilaments }
\author{Herv\'{e} Mohrbach$^{1,2}$ and Igor M. Kuli\'{c}$^{2}$}
\date{\today}

\begin{abstract}
We investigate an analytically tractable toy model for thermally induced
polymorphic dynamics of cooperatively rearranging biofilaments - like
microtubules. The proposed 4 -block model, which can be seen as a
coarse-grained approximation of the full polymorphic tube model, permits a
complete analytical treatment of all thermodynamic properties including
correlation functions and angular fourier mode distributions. Due to its
mathematical tractability the model straightforwardly leads to some physical
insights in recently discussed phenomena like the "length dependent
persistence length". We show that a polymorphic filament can disguise itself
as a classical wormlike chain on small and on large scales and yet display
distinct anomalous tell-tale features indicating an inner switching dynamics
on intermediate length scales.

\end{abstract}
\affiliation{$^{1}$Groupe BioPhysStat, Universit\'{e} Paul Verlaine-Metz, 57078 Metz,
France }
\affiliation{$^{2}$ CNRS, Institut Charles Sadron, 23 rue du Loess BP 84047, 67034
Strasbourg, France }

\pacs{87.16.Ka, 82.35.Pq, 87.15.-v}
\maketitle

\section{Introduction}

Biological filaments of the cytoskeleton are large macromolecules
formed by a hierarchical self-assembly of smaller yet often highly
complex protein subunits. The monomer complexity can allow the
subunits to undergo rearrangements between several conformational
states. Once inserted into a macromolecular lattice these
individual subunits can start to interact giving rise to
cooperative phenomena which can affect equilibrium and dynamical
properties of the whole assembly in unexpected manners. It is the
goal of this paper to explore in detail the statistical mechanics
of this type of cooperatively switching supramolecular assemblies,
whose paradigm example might be found in the stiffest cytoskeletal
protein filaments of eukaryotic cells, the microtubules
\cite{Alberts}. Microtubules are hollow nanotubes whose walls are
formed by lateral self-association of parallel protofilaments that
themselves are built by a head-to-tail polymerization of
$\alpha\beta$-tubulin heterodimer protein subunits. This very
complex architecture confers to microtubules their high stiffness
as well as a number of unique static and dynamic properties. In
\cite{MTKulic1}, guided by experimental findings, we built the
case for a novel model of microtubules, based on two hypotheses:
the elementary tubulin dimer units can fluctuate between a curved
and a straight configuration and can interact cooperatively along
a protofilament axis. This implies that the ground state of
microtubules is not, as usually accepted a straight Euler beam,
but instead a fluctuating several micron sized cooperative
super-helix. The resulting polymorphic dynamics of the microtubule
lattice seems to quantitatively explain several experimental
puzzles including anomalous scaling of dynamic fluctuations of
grafted microtubules \cite{Taute}, their apparent length-stiffness
relation \cite{Pampaloni} and their remarkably curved-helical
appearance in general \cite{Venier}. These results rely on
phenomenological modelling where the cylindrical symmetry of the
microtubule lattice is approximated by a continuous symmetry
\cite{MTKulic1}. This approximation seems reasonable as the number
of protofilaments is typically large, ranging from $9$ to $17$ for
taxol stabilized in-vitro microtubules with a predominant $14$
protofilament structure \cite{Wade,Chretien}. In vivo,
microtubules most commonly appear with $13$ protofilaments\
\cite{Bouchet}, although many exceptions exist depending of the
cell type. The approach developed in \cite{MTKulic1} revealed the
existence of a unique and unusual zero mode dynamics which has
strong consequences : microtubules are permanently coherently
reshaping -i.e. changing their reference ground state
configuration- by thermal fluctuations.

Going beyond the continuous phenomenological approach of
\cite{MTKulic1}, we explore here other important aspects of the
polymorphic microtubules theory by considering a simplified
minimal solvable model. We will adopt a coarse-grained approach
where the microtubule lattice is considered as made up of only $4$
blocks of protofilaments that can fluctuate between a curved and a
straight configuration. Although the coarse-grained model to some
extent loses the (quasi)continuous zero mode dynamics, it captures
a number of important features of the full model \cite{MTKulic1}
and exact analytic computations of relevant observables becomes
accessible in this simpler case. In particular the tubulin-tubulin
state correlation function, the persistence length, the
thermodynamic stability of conformational states as well as the
tangent angle spectrum can be computed rather elegantly. This
simplified approach should give more detailed analytical insights
into microtubule's static properties and make connection with the
previous phenomenological model's parameters introduced in
\cite{MTKulic1}. While it is not our prime goal here to compare
the developed toy model with experiments we hope that some of the
derived results (like the tangent angle power spectrum) can become
useful guides for future experimental quantification of
polymorphic filament fluctuations. For a more detailed
experiment-theory comparison and deeper motivation of the model we
refer the reader to the articles \cite{MTKulic1}.

\section{The 4-block polymorphic tube model}

The microtubule lattice (see Fig. 1) is modelled as a continuum tube material
made of a variable number of protofilaments with $R_{i}\approx7.5nm,$
$R_{o}\approx11.5nm$ - the inner and outer microtubule radii respectively. The
protofilaments are twisted around the microtubule's longitudinal axis with the
corresponding internal twist $q_{0}$ - or equivalently the pitch $\lambda=2\pi
q_{0}^{-1}$ - being a lattice type dependent constant that takes typical
discrete values $\lambda=+3.4\mu m,+25\mu m,-6\mu m$ for $12$, $13$ and $14$
protofilament microtubules respectively \cite{Wade,Chretien}\cite{Ray}%
\cite{ChretienFuller}.

\begin{figure}[ptbh]
\begin{center}
\includegraphics[
width=3.1081in ]{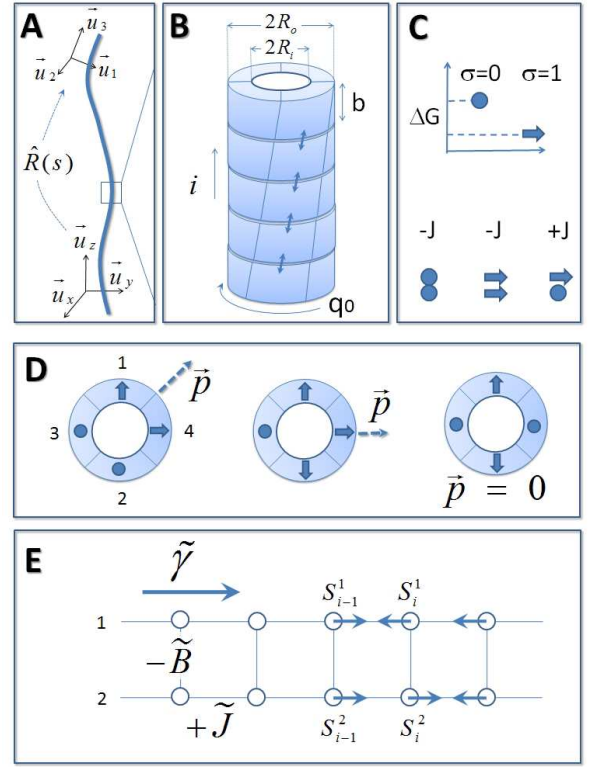}
\end{center}
\caption{Elements of the "polymorphic four block tube model". (a) Tube contour
with the internal and external frames. (b) A short tube section with intrinsic
twist $q_{0}$ with four blocks at each section. (c) The blocks can fluctuate
cooperatively between two discrete states. The outwards curved $\sigma=1$
(arrow) state is energetically preferred over the straight state $\sigma=0$
(circle) with an energy gain $E=-\Delta G$. The junction between straight and
curved states along the longitudinal block axis are penalized by a coupling
constant $E=+J.$ (d) Competition between block switching energy and elastic
lattice strain energy leads to spontaneous symmetry breaking: the tube bends
to a randomly chosen direction and assumes a non-zero vectorial polymorphic
order parameter $\vec{p}$ ("conformational polarization") (e) The four block
model can be decoupled into a pair of two-block models which can be mapped
onto a ladder type Ising model with coupling constants $-\widetilde{B}$ and
$+\widetilde{J}$ and a bias field $\widetilde{\gamma}$.}%
\end{figure}To describe the tube's geometry we introduce two reference frames
(cf. Fig. 1A). One is the material frame with base vectors $(\vec{u}_{1}%
,\vec{u}_{2},\vec{u}_{3})$ attached to each microtubule cross-section. The
other is an external fixed laboratory frame with base vectors $(\vec{u}%
_{x},\vec{u}_{y},\vec{u}_{z})$ with respect to which all deformations are
measured. Putting the filament along the $\vec{u}_{z}$ axis direction and
considering small angular deflections we have $\vec{u}_{z}\approx\vec{u}_{3}$.
In this case the two frames are simply related by a rotation transformation
$\hat{R}(s)$ given by internal microtubule lattice twist $q_{0}$, such that
$(\vec{u}_{x},\vec{u}_{y})=\hat{R}(s)(\vec{u}_{1},\vec{u}_{2})$ with
\begin{equation}
\hat{R}(s)=\left(
\begin{array}
[c]{cc}%
\cos(q_{0}s+\varphi) & -\sin(q_{0}s+\varphi)\\
\sin(q_{0}s+\varphi) & \cos(q_{0}s+\varphi)
\end{array}
\right)  \label{R}%
\end{equation}
and $s\in\left[  0,L\right]  $ the longitudinal position variable along the
microtubule centerline. The arbitrary rotation angle $\varphi$ corresponds to
the angular deviation between the two frames at $s=0.$ As each protofilament
is built by a self association of $N$ discrete GDP-tubulin dimers of length
$b\approx8nm$ it is natural to introduce a discrete variable $i$ such that all
microtubule's cross-section positions are written as $s=ib$ with $i=1..N.$

The coarse-grained block approximation consists of grouping neighboring
protofilaments into $n=4$ effective blocks (cf. Fig. 1B). The fluctuations of
the block-dimers between 2 states - a straight and a curved state with
intrinsic curvature $\kappa_{FP}$ and with an energy difference $\Delta G>0$
favoring the curved state are modelled by a two state variable $\sigma_{i}%
^{k}=0,1$ where $i=1..N$ is the longitudinal position and $k=1..4$ the block's
index (cf. Fig. 1C).

To complete the description of the polymorphic tube model two order parameters
at each microtubule cross-section can be introduced \cite{MTKulic1}. The first
is the \textit{vectorial polymorphic order parameter} $\vec{p}_{i}=\vec{u}%
_{1}(\sigma_{i}^{1}-\sigma_{i}^{2})+\vec{u}_{2}(\sigma_{i}^{4}-\sigma_{i}%
^{3}),$ a 2D vector attached to each cross-section (cf Fig. 1D) describing the
asymmetry of the curved state distribution - a kind of "conformational
polarization vector" of the block states. For instance the "all-straight" or
"all-curved" protofilament state correspond both to the same value $\vec{p}%
=0$, cf. Fig. 1D (as the curved state distribution is isotropic in both
special cases). A second (scalar) quantity $m_{i}=\sum_{k=1}^{4}\sigma_{i}%
^{k}$ counts the total number of blocks in the curved state at each
cross-section $i$. The tubulin cooperativity \textit{along} each protofilament
(block) axis is modelled by an Ising type nearest-neighbor cooperative
interaction with an interaction energy $J>0$ favoring longitudinal nearest
neighbors to be in the same state. The interaction energy at cross-section $i$ reads:%

\begin{equation}
e_{inter,i}=-J\sum\nolimits_{k=1}^{4}\left(  2\sigma_{i}^{k}-1\right)  \left(
2\sigma_{i+1}^{k}-1\right)  \label{eJ}%
\end{equation}
The total elastic +\ polymorphic energy can then be written as \cite{MTKulic1}%
:
\begin{equation}
E_{MT}=\sum\nolimits_{i=1}^{N}\left(  e_{i}+e_{inter,i}\right)  .\label{EMT}%
\end{equation}
where $e_{i}$ is the combined elastic energy and the energy resulting from the
switching of tubulin dimers at i-th cross-section:%
\begin{equation}
e_{i}=\frac{Bb}{2}\left[  \left(  \vec{\kappa}_{i}-\vec{\kappa}_{pol,i}%
\right)  ^{2}+\frac{\kappa_{1}^{2}}{2}\left(  \frac{\pi\gamma}{2}m_{i}%
-p_{i}^{2}\right)  \right]  \label{e}%
\end{equation}
with the elastic bending modulus $B=\frac{Y\pi}{4}\left(  R_{o}^{4}-R_{i}%
^{4}\right)  $ and $Y$ the Young modulus. The effective lattice curvature
$\kappa_{1}=\frac{\left(  R_{0}-R_{1}\right)  ^{2}}{\pi\left(  R_{o}^{2}%
+R_{i}^{2}\right)  }\kappa_{FP}$ results directly from the preferred curvature
of the individual protofilament $\kappa_{PF}$ \cite{MTKulic1}. Here
$\vec{\kappa}$ is the microtubule centerline curvature vector and $\vec
{\kappa}_{pol}$ the polymorphic curvature vector which in the external
coordinate frame $\left(  \vec{u}_{x},\vec{u}_{y}\right)  $ is
\begin{equation}
\vec{\kappa}_{pol,i}=2^{-1/2}\kappa_{1}\hat{R}(ib)\vec{p}_{i}\label{kapppol}%
\end{equation}
In Eq. \ref{e} we introduced an important dimensionless parameter
$\gamma=\frac{\kappa_{PF}}{\kappa_{1}}-\frac{8\Delta G}{bB\kappa_{1}^{2}}$
which measures the competition between block switching and elastic energy and
ultimately determines the microtubule shape. It effectively acts as an
external field that biases the curved lattice state for $\gamma<0$ and favors
the straight state for $\gamma>0$ (cf. below). For small deflections
$\theta_{x/y}<<1$ around the $z$-axis, the unit vector tangent to the
microtubule's centerline is approximately given by $\vec{t}\approx(\theta
_{x},\theta_{y},1)$\ in the laboratory frame $\left(  \vec{u}_{x},\vec{u}%
_{y},\vec{u}_{z}\right)  $\ where $\vec{\theta}=\left(  \theta_{x},\theta
_{y}\right)  $\ are the centerline deflection angles in x/y direction. The
global centerline curvature vector $\vec{\kappa}=d\vec{t}/ds$\ can then be
approximated as $\overrightarrow{\kappa}\approx\left(  \theta_{x}^{\prime
},\theta_{y}^{\prime},0\right)  .$ Writing the total curvature as
$\overrightarrow{\kappa}=\overrightarrow{\kappa}_{pol}+\overrightarrow
{\theta^{\prime}}_{el}$ with $\overrightarrow{\theta^{\prime}}_{el}$ the
purely elastic contribution we readily see from Eqs. \ref{eJ},\ref{EMT} that
the partition function decomposes into a product of two independent elastic
and polymorphic contributions $Z=Z_{el}Z_{pol},$ with $Z_{el}=\int D\theta
\exp\left(  -\frac{l_{B}}{2}\int\nolimits_{0}^{L}\theta_{el}^{\prime
2}ds\right)  $ (and $l_{B}=B/k_{B}T$ the bending persistence length with
$k_{B}T$ the thermal energy). Therefore our main goal reduces to the
computation of $Z_{pol}=\sum_{\sigma}\exp(-\frac{1}{k_{B}T}E_{\sigma})$ with
$E_{\sigma}$ the polymorphic contribution of $E_{MT}$. In the general case
(with large number of blocks) this appears to be a formidable task, that
however can be exactly performed in the case of $4$ blocks as we will see in
the following.

\subsection{Reduction to a 2 $\times$ 2-block model}

We first note that, at each cross-section $p_{i}^{2}=\left(  \sigma_{i}%
^{4}-\sigma_{i}^{3}\right)  ^{2}+(\sigma_{i}^{2}-\sigma_{i}^{1})^{2}$
decomposes into a sum over mutually facing blocks i.e. only blocks 1-2 and 3-4
couple directly. Consequently the partition function can be factorized into a
product of two simpler ones $Z_{pol}=Z_{12}Z_{34}$ with $Z_{kl}$ the partition
function of the Ising variables $\sigma^{k}$ and $\sigma^{l}$ of the blocks
that face each other, cf. Fig. 1D. Therefore from now on we may consider the
$2$\ block model, say the block pair $1-2$. The polymorphic order parameters
are in this case $m_{i}=\sigma_{i}^{1}+\sigma_{i}^{2}$ and $\overrightarrow
{p}_{i}=\left(  \sigma_{i}^{1}-\sigma_{i}^{2}\right)  \vec{u}_{1}.$ By
introducing new more convenient Ising variables $S_{i}^{k}=\pm1$ defined as
$S_{i}^{k}=2\sigma_{i}^{k}-1$ the energy Eq. \ref{eJ} takes the familiar form
of a ladder type Ising model (cf. Fig. 1E). In terms of these variables,
$Z_{12}=Z_{0}\widetilde{Z}$ with $Z_{0}=\exp(\frac{1}{k_{B}T}B\kappa_{1}%
^{2}L\left(  \pi\gamma-1\right)  /4)$ and
\begin{equation}
\widetilde{Z}=\sum_{\left\{  S^{1},S^{2}\right\}  =-1,1}\exp(-\widetilde
{E})\label{Ztilde}%
\end{equation}
where from Eqs. \ref{eJ}-\ref{e} we obtain
\begin{equation}
\widetilde{E}= {\textstyle\sum_{i=1}^{N}}
\left[  \widetilde{B}S_{i}^{1}S_{i}^{2}+\widetilde{\gamma}\left(  S_{i}%
^{1}+S_{i}^{2}\right)  -\widetilde{J}\left(  S_{i}^{1}S_{i+1}^{1}+S_{i}%
^{2}S_{i+1}^{2}\right)  \right]  \label{E12}%
\end{equation}
with $\widetilde{B}=\frac{B\kappa_{1}^{2}b}{4k_{B}T},$ $\widetilde{J}=\frac
{J}{k_{B}T}$ and $\widetilde{\gamma}=\frac{\pi}{2}\gamma\widetilde{B}.$ With
the two spins - on either side of the ladder (Fig. 1E)- being interchangeable
we have $\left\langle S_{i}^{1}\right\rangle =\left\langle S_{i}%
^{2}\right\rangle $ (and thus $\left\langle \sigma_{i}^{1}\right\rangle
=\left\langle \sigma_{i}^{2}\right\rangle $) for all $i$ as the energy is
translationally invariant. From Eq. \ref{kapppol}, we readily see that
$\left\langle \vec{\kappa}_{pol,i}\right\rangle =0.$ And yet this does not
mean that the microtubule is on average in a straight configuration. Just on
the contrary, the microtubule can form a three dimensional super helix
\cite{MTKulic1} that is "coherently" reshaping as the curvature alternately
exchanges from one block to the other with a non-vanishing value of
$\left\langle \kappa_{pol}^{2}\right\rangle $. Therefore the order parameter
that characterizes the typical curvature of the lattice is the mean
"polarization" $P=\sqrt{N^{-1}\sum_{i}\left\langle \overrightarrow{p}_{i}%
^{2}\right\rangle }$ of the curved states (as $\left\langle \kappa_{pol}%
^{2}\right\rangle =\kappa_{1}^{2}P^{2}$). The other important
quantity is $M=\frac{1}{N}\sum\nolimits_{i}\left\langle
\sigma_{i}^{k}\right\rangle =\left\langle
\sigma_{i}^{k}\right\rangle $ - the mean concentration of curved
states, or in the Ising terminology the mean "magnetization" (up
to a trivial additive constant) . Within the same terminology the
parameter $\widetilde {\gamma}$ takes the role of a "magnetic
field" that, according to its sign, favors one or the other
possible spin orientation (curved or straight block) -but the same
orientation for blocks on both sides of the ladder in Fig. 1E. The
"ferromagnetic coupling constant " $\widetilde{J}$ promotes the
longitudinal parallel alignments of spins along a given block
axis. The "anti-ferromagnetic coupling" $\widetilde{B}$ on the
other hand favors $S_{i}^{1}$ and $S_{i}^{2}$ to be antiparallel -
a tendency which competes with $\widetilde{\gamma}$ and the
lattice is frustrated. For large $\left\vert
\widetilde{\gamma}\right\vert >>\widetilde{B}$, we therefore
expect the alignment tendency to win on both sides of the ladder
so that $\left\langle S_{i}^{1/2}\right\rangle \approx\pm1,$ ($+1$
if $\widetilde{\gamma}<0$ and $-1$ if $\widetilde{\gamma}>0$) and
thus $M\approx1$ or $0.$ In this situation the microtubule is in a
straight state with $P\approx0$, which is either completely
unstressed with $M\approx$ $0$ (for positive $\widetilde{\gamma}$)
or maximally prestrained state with $M\approx1$ (for negative
$\widetilde {\gamma}$).

For $\widetilde{\gamma}=0,$ the lattice is not frustrated and on average when
on one ladder-side $S_{i}^{1}=1$ then $S_{i}^{2}=-1$ on the other (and vice
versa). Consequently $\left\langle S_{i}^{k}\right\rangle =0$ and
$M=\left\langle \sigma_{i}^{k}\right\rangle =1/2.$ In this case blocks $1$ and
$2$ at the cross-section $i$ fluctuate alternately between the curved and
straight state and the tube bends alternately in the directions $\vec{u}_{1}$
and $-\vec{u}_{1}.$ The cooperative ferromagnetic interaction $\widetilde{J}$
implies a correlation between the spins along the contour and the formation of
domains with size of the order of the spin-spin correlation length $\xi$
computed below. It is these whole domains that alternately bend the tube in
the $\vec{u}_{1}$ and $-\vec{u}_{1}$ direction that lead to $\left\langle
\overrightarrow{p}_{i}\right\rangle =0$ but $P\neq0.$ In fact $P$ will take
its largest value $P\approx1$ in this non frustrated case. One remarks that
for $\xi$ much smaller that the internal twist wavelength $\lambda$ the
typical domain looks like a circular arc, whereas for $\xi\gg\lambda,$ a
typical coherent domain is a super-helix in the 3 dimensional space with pitch
$\lambda$ as the direction of bending $\pm\vec{u}_{1}$ is slowly rotating in
the external frame. For $L>>\xi,$ the microtubule is made up of a
juxtaposition of uncorrelated fluctuating helices or of uncorrelated circular
arcs that bend independently in the $\pm\vec{u}_{1}$ direction. Consequently
we expect a long distance behavior similar to an usual worm like chain. This
is true despite the fact that elastic contributions where neglected in this
qualitative discussion. Indeed there are no elastic fluctuations here (as for
usual semiflexible filaments): the polymorphic transition from curved to
straight states of short uncorrelated segments only mimic elastic
fluctuations. For $\widetilde{\gamma}$ small but non zero the lattice is
slightly frustrated and thus less curved (smaller value of $P$) displaying
nevertheless a qualitatively similar physical behavior.

\section{2-block model thermodynamics}

The partition function Eq. \ref{Ztilde}, can be exactly computed via the
transfer matrix method. For this purpose, we impose periodic boundary
conditions on both sides of the ladder $S_{1}^{1/2}=S_{N+1}^{1/2}$. This
permits us to write the partition function as $\widetilde{Z}=$Tr$(T^{N})$ with
the symmetric $\left(  4\times4\right)  $ matrix $T$ defined by
\begin{align*}
&  \left\langle S_{i}^{1},S_{i}^{2}\right\vert T\left\vert S_{i+1}^{1}%
,S_{i+1}^{2}\right\rangle =\\
&  e^{-\frac{\widetilde{\gamma}}{2}(S_{i}^{1}+S_{i}^{2}+S_{i+1}^{1}%
+S_{i+1}^{2})-\frac{\widetilde{B}}{2}\left(  S_{i}^{1}S_{i}^{2}+S_{i+1}%
^{1}S_{i+1}^{2}\right)  +\widetilde{J}S_{i}^{1}S_{i+1}^{1}+\widetilde{J}%
S_{i}^{2}S_{i+1}^{2}}%
\end{align*}
with explicit elements given by
\[
T=\left(
\begin{array}
[c]{cccc}%
e^{2\widetilde{J}-2\widetilde{\gamma}-\widetilde{B}} & e^{-\widetilde{\gamma}}
& e^{-\widetilde{\gamma}} & e^{-2\widetilde{J}-\widetilde{B}}\\
e^{-\widetilde{\gamma}} & e^{2\widetilde{J}+\widetilde{B}} & e^{-2\widetilde
{J}+\widetilde{B}} & e^{\widetilde{\gamma}}\\
e^{-\widetilde{\gamma}} & e^{-2\widetilde{J}+\widetilde{B}} & e^{2\widetilde
{J}+\widetilde{B}} & e^{\widetilde{\gamma}}\\
e^{-2\widetilde{J}-\widetilde{B}} & e^{\widetilde{\gamma}} & e^{\widetilde
{\gamma}} & e^{2\widetilde{J}+2\widetilde{\gamma}-\widetilde{B}}%
\end{array}
\right)
\]
Denoting $U$ the matrix diagonalizing $T$, such that $UTU^{-1}=\Lambda$ is a
diagonal matrix with eigenvalues $\lambda_{r}$ with $r=1,4$ , the partition
function can be written as $Z=Tr\left(  \Lambda^{N}\right)  =\sum
_{r=1,4}\lambda_{r}^{N}$. Denoting $\lambda_{1}$ the largest eigenvalue the
free energy per lattice site $f=F/N=-k_{B}T/N\ln Z$ reduces in the
thermodynamic limit of $N\rightarrow\infty$ to the expression $f=-k_{B}%
T\ln\lambda_{1}.$ From the free energy, all other thermodynamic quantities can
be derived. In particular the curved state density $M=\frac{1}{2}-\frac{1}%
{4}\frac{\partial}{\partial\widetilde{\gamma}}\ln\lambda_{1}$ and the
polarization $P=\sqrt{\frac{1}{2}+\frac{1}{2}\frac{\partial}{\partial
\widetilde{B}}\ln\lambda_{1}}.$

The explicit expression for $\lambda_{1}$ is rather cumbersome and we omit it
here. Instead we provide the plot of $M$ and $P$ in terms of $\widetilde
{\gamma}$ for different values of $\widetilde{J}$ with $\widetilde{B}=1$ (in
\cite{MTKulic1}, $\widetilde{B}$ was indeed found to be of order unity) in
Fig. 2. We first remark that $P(-\gamma)=P(\gamma)$, so the microtubule's mean
curvature is symmetric with respect to the sign of $\gamma.$ Further we
observe that there always exists a range of $\widetilde{\gamma}$ where $P\ $is
close to unity which is quite spread for small coupling $J\ll k_{B}T$ and that
for $\left\vert \widetilde{\gamma}\right\vert \gg\widetilde{B},$ the
polarization $P$ is larger for smaller values of $J.$ Although in this regime
the total energy density is minimal for straight unstressed or prestrained
states, these states have a small entropy. In comparison, curved tube states
have a higher energy density in this regime but also a higher entropy, as
blocks of small size $\xi$ fluctuate independently. Consequently these curved
-helical- states can have a smaller free energy. With growing coupling $J$,
the size of coherent blocks increases (as $\xi$ grows with $J$) and the energy
contribution becomes dominant over entropy so $P$ gets lower with growing $J.$
From Fig. 2, we observe that for $J=1.5k_{B}T,$ the entropy is already
negligible as for $\left\vert \widetilde{\gamma}\right\vert >\widetilde{B}$
the straight states with $P\approx0$ are selected. On the contrary for
$\left\vert \widetilde{\gamma}\right\vert <\widetilde{B},$ the
microtubule\ adopts a curved conformation with a quasi constant value
$P\approx1$- corresponding to maximum lattice curvature which is independent
of $\widetilde{\gamma}$. This plateau region shows that in order to observe a
curved (or helical) conformation microtubule does not require a very precise
fine tuning of $\widetilde{\gamma}$ - as long it is smaller than
$\widetilde{B}$. This approximate $\widetilde{\gamma}$ independence allows us
to limit ourselves to the analytically most elegant case $\widetilde{\gamma
}=0$. \begin{figure}[ptbh]
\begin{center}
\includegraphics[
width=3.1081in ]{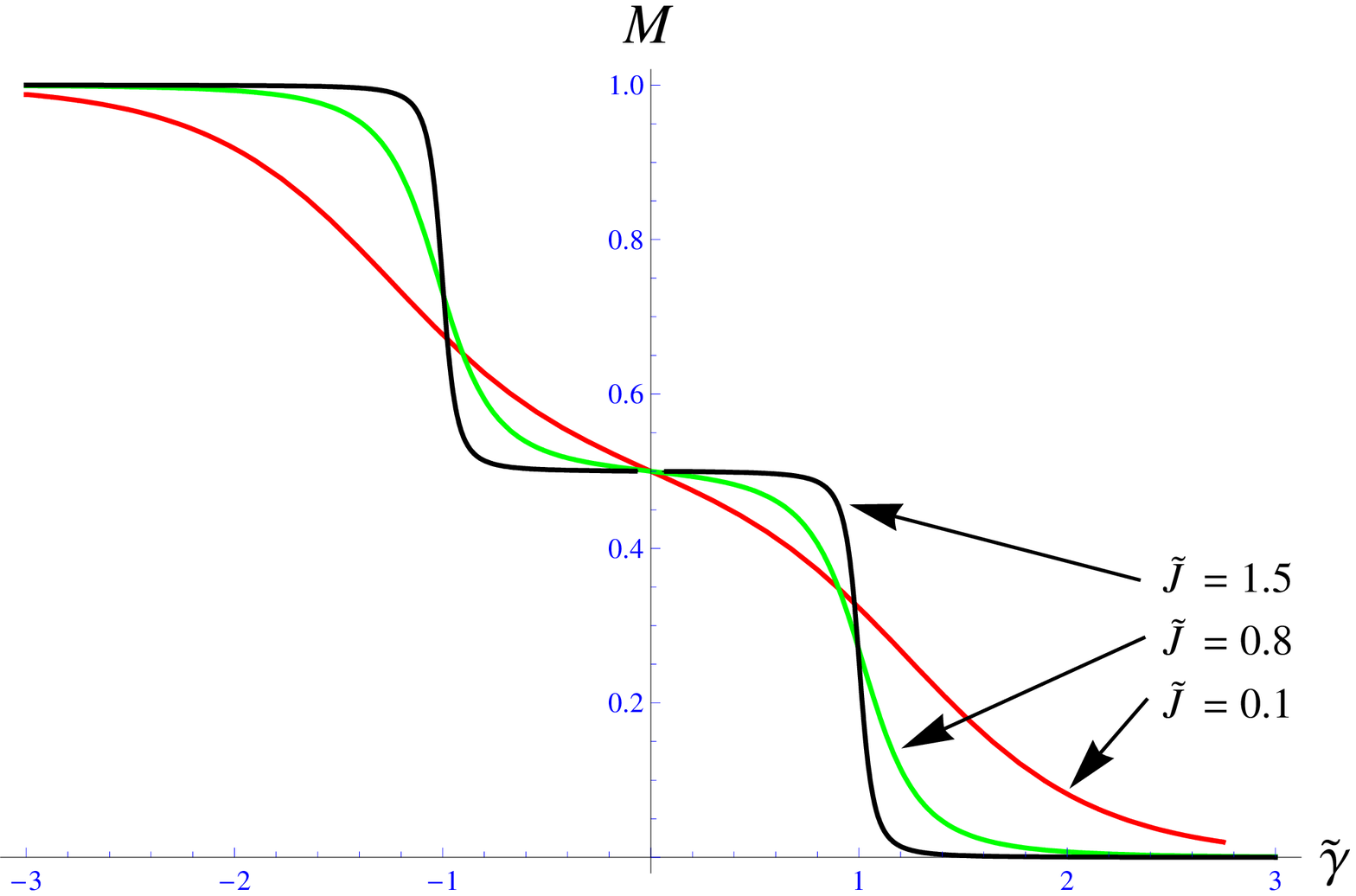}
\end{center}
\caption{The curved state density $M=\frac{1}{N}\sum\nolimits_{i=1}%
^{N}\left\langle \sigma_{i}^{k}\right\rangle $ versus the state bias parameter
$\widetilde{\gamma}$ for different values of coupling constant$\ \widetilde
{J}$ and fixed stiffness $\widetilde{B}=1.$ Note the broad $\widetilde{\gamma
}$ independent plateau region for $\widetilde{J}=1.5$ where $M=1/2.$ For large
values of $\left\vert \widetilde{\gamma}\right\vert ,$ $M$ is either $0$ or
$1$ corresponding to straight unstrained or straight strained lattice states
respectively. }%
\end{figure}\begin{figure}[ptbh]
\begin{center}
\includegraphics[
width=3.1081in ]{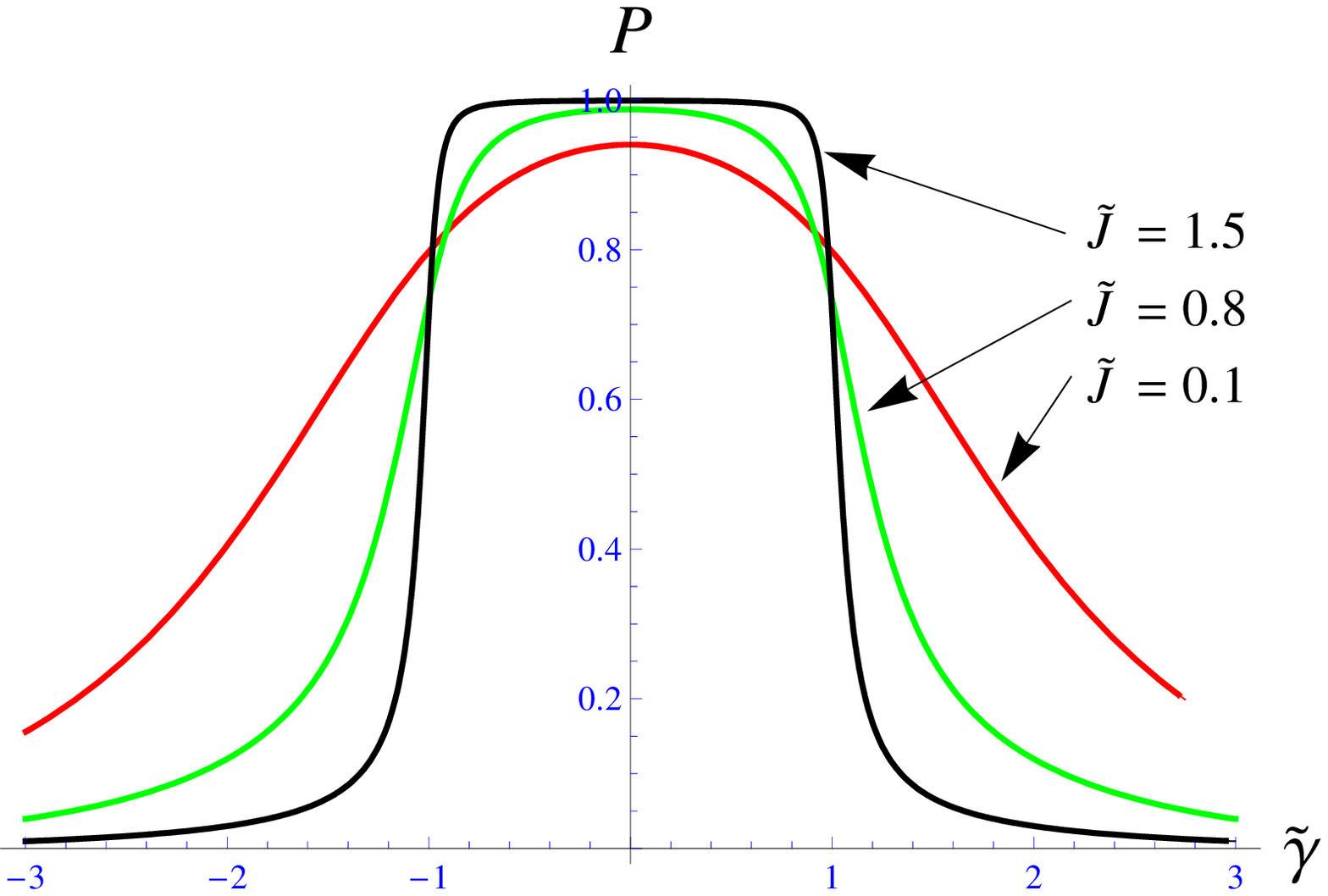}
\end{center}
\caption{Plot of the mean polymorphic polarization $P$ of the reduced two
block model versus the conformational bias field $\widetilde{\gamma}$ for
different values of coupling constant $\widetilde{J}$ and effective tube
stiffness $\widetilde{B}=1.$ For $\widetilde{J}\geq1.5,$ the tube's curvature
is at maximum for $\left\vert \widetilde{\gamma}\right\vert <1,$ as
$P\approx1$ in this region. For larger $\left\vert \widetilde{\gamma
}\right\vert ,$ the tube is in a straight state and $P$ goes to zero. }%
\end{figure}

\subsection{Simple case $\widetilde{\gamma}=0$}

In this case all quantities can be computed in compact form. In particular the
largest eigenvalue of the transfer matrix simplifies to $\lambda_{1}%
=2\cosh(\widetilde{B})\cosh(2\widetilde{J})+D$ with $D=(3+\cosh(2\widetilde
{B})+2\cosh(4\widetilde{J})\sinh^{2}(\widetilde{B}))^{1/2}.$ As already
mentioned the curved state density is constant $M=1/2$ (independent of
$\widetilde{B}$ and $\widetilde{J}$), whereas the total "polarization" is
$P^{2}=1-2\left\langle \sigma_{i}^{1}\sigma_{i}^{2}\right\rangle =\frac{1}%
{2}+\frac{\sinh(\widetilde{B})\cosh(2\widetilde{J})}{D}$ which tends to $1$
for $\widetilde{J}\gtrsim1$. This means that $\left\langle \sigma_{i}%
^{1}\sigma_{i}^{2}\right\rangle \rightarrow0$ which can be understood as both
states (on opposing ladder sides) at the same site $i$ fluctuate alternately
between $0$ and $1.$ The coupling $\widetilde{J}$ which is a measure of the
cooperativity determines over which distance the curved state is coherent. The
cooperativity can be measured by the spin-spin correlation function
$g_{ij}^{kl}=\left\langle \sigma_{i}^{k}\sigma_{j}^{l}\right\rangle
-\left\langle \sigma_{i}^{k}\right\rangle \left\langle \sigma_{j}%
^{l}\right\rangle =\left\langle \sigma_{i}^{k}\sigma_{j}^{l}\right\rangle
-1/4$ with $k,l=1,2$ at two different sites $i$ and $j$. By symmetry
$g_{ij}^{11}=g_{ij}^{22}$ and $g_{ij}^{12}=g_{ij}^{21}.$ Introducing the two
matrices $\underline{V}^{1}$ and $\underline{V}^{2}$ defined as
\[
\underline{V}^{1}=\left(
\begin{array}
[c]{cccc}%
1 & 0 & 0 & 0\\
0 & 1 & 0 & 0\\
0 & 0 & -1 & 0\\
0 & 0 & 0 & -1
\end{array}
\right)  \text{ \ and }\underline{V}^{2}=\left(
\begin{array}
[c]{cccc}%
1 & 0 & 0 & 0\\
0 & -1 & 0 & 0\\
0 & 0 & 1 & 0\\
0 & 0 & 0 & -1
\end{array}
\right)
\]
the correlation functions of the variable $S^{k}$ with $k=1,2$ can be written
as $\left\langle S_{i}^{k}S_{j}^{l}\right\rangle =\frac{1}{Z}$Tr$(T^{i-1}%
\underline{V}^{k}T^{j-i-1}\underline{V}^{l}T^{N-j+i+1})$ for $i<j.$ With the
matrix $U$ diagonalizing $T$, i.e. $U^{-1}TU=\Lambda$ and the following
transformed matrices $V^{k}=U^{-1}\underline{V}^{k}U$, we have $\left\langle
S_{i}^{k}S_{j}^{l}\right\rangle =\frac{1}{Z}$Tr$(\Lambda^{i-1}V^{k}%
\Lambda^{j-i}V^{l}\Lambda^{N-j+1}).$ In the limit $N\rightarrow\infty,$ this
expression can be evaluated :%

\begin{align}
\left\langle S_{i}^{k}S_{j}^{l}\right\rangle  &  =V_{11}^{k}V_{11}^{l}+\left(
\frac{\lambda_{2}}{\lambda_{1}}\right)  ^{j-i}V_{21}^{k}V_{12}^{l}\nonumber\\
&  +\left(  \frac{\lambda_{3}}{\lambda_{1}}\right)  ^{j-i}V_{31}^{k}V_{13}%
^{l}+\left(  \frac{\lambda_{4}}{\lambda_{1}}\right)  ^{j-i}V_{41}^{k}%
V_{41}^{l} \label{SpinSpin}%
\end{align}
where the eigenvalues are ordered $\lambda_{1}>..>\lambda_{4}.$ From Eq.
\ref{SpinSpin} all $g_{ij}^{kl}$ can be deduced. It turns out that the most
interesting quantity connected to the spatial fluctuations of the microtubule
-its persistence length (see next section) - is connected to $\Delta
g_{ij}\equiv g_{ij}^{11}-g_{ij}^{12}=\left\langle \sigma_{i}^{1}\sigma_{j}%
^{1}\right\rangle -\left\langle \sigma_{i}^{1}\sigma_{j}^{2}\right\rangle .$
An explicit computation of the matrix elements of $V^{k}$ shows that
$V_{14}^{k}=V_{11}^{k}=0$ and $V_{31}^{1}V_{13}^{1}=V_{13}^{2}V_{31}^{1}$ as
well as $V_{21}^{1}V_{12}^{1}=-V_{12}^{2}V_{21}^{1}=P^{2}$. This leads to
\begin{equation}
\Delta g_{ij}=\frac{1}{2}P^{2}\exp(-\left\vert i-j\right\vert /\widetilde{\xi
}) \label{CC}%
\end{equation}
where the correlation length is $\widetilde{\xi}=\left[  \ln\left(
\frac{\lambda_{1}}{\lambda_{2}}\right)  \right]  ^{-1}$ with $\lambda
_{2}=2\exp(\widetilde{B})\sinh(2\widetilde{J}).$ With the result for $2$
blocks (ladder model, Fig 1E) we are now able to compute the persistence
length for the $4$-block polymorphic tube (Fig 1B).

\section{Angular persistence length}

The persistence length $l_{p}$ is the length scale characterizing the
filament's resistance to thermally induced bending moments. For usual
biofilaments like DNA or actin filaments $l_{p}=l_{B}$ is a material constant
equal to the bending persistence length $l_{B}\equiv B/k_{B}T$. However for
microtubules $l_{p}$ is known to be length dependent \cite{Pampaloni}. Here we
compute the angular persistence length for the $4$-block model from the usual
definition in terms of the tangent-tangent correlation function $\left\langle
\overrightarrow{t}(s^{\prime})\cdot\overrightarrow{t}(s)\right\rangle
=\exp(-\left\vert s^{\prime}-s\right\vert /l_{p})$ where $\overrightarrow
{t}(s)$ is the unit-tangent vector at the position $s$ of the microtubule
centerline. From $\overrightarrow{t}(s)=(\theta_{x}(s),\theta_{y}%
(s),\sqrt{1-\theta_{x}^{2}(s)-\theta_{y}^{2}(s)})$ in the external frame, we
deduce to quadratic order in $\theta$ that $\left\langle \overrightarrow
{t}(s^{\prime})\cdot\overrightarrow{t}(s)\right\rangle \approx1-\frac{1}%
{2}V(s^{\prime}-s)$ with the angular variance $V(s^{\prime}-s)=\left\langle
(\theta_{x}(s^{\prime})-\theta_{x}(s))^{2}\right\rangle +\left\langle
(\theta_{y}(s^{\prime})-\theta_{y}(s))^{2}\right\rangle .$ Therefore one can
write $\left\langle \overrightarrow{t}(s^{\prime})\cdot\overrightarrow
{t}(s)\right\rangle \approx\exp(-\left\vert s^{\prime}-s\right\vert /l_{p})$
with a persistence length $l_{p}(s^{\prime}-s)=2\frac{\left\vert s^{\prime
}-s\right\vert }{V(s^{\prime}-s)}$ that will be manifestly distance dependent
in our case. Now by writing $\overrightarrow{\theta}=\overrightarrow{\theta
}_{pol}+\overrightarrow{\theta}_{el}$ and from the independence of polymorphic
and elastic contributions $\left\langle \overrightarrow{\theta}_{pol}%
\cdot\overrightarrow{\theta}_{el}\right\rangle =\left\langle \overrightarrow
{\theta}_{pol}\right\rangle \cdot\left\langle \overrightarrow{\theta}%
_{el}\right\rangle =0,$ we deduce that the total persistence length can be
decomposed as $l_{p}=(\frac{1}{l_{B}}+\frac{1}{l_{pol}})^{-1}$ with a
polymorphic persistence length
\begin{equation}
l_{pol}(s^{\prime}-s)=2\frac{\left\vert s^{\prime}-s\right\vert }%
{V_{pol}(s^{\prime}-s)}\label{PL}%
\end{equation}
At this level, for the geometric description of the microtubule shape in space
it is more convenient to use a continuum description and replace the discrete
index $i$ by the continuous variable $s,$ so that $\overrightarrow{\theta
}_{pol,i}\rightarrow\overrightarrow{\theta}_{pol}(s)$ which is obtained from
the integration over $s$ of the curvature Eq. \ref{kapppol}
\begin{equation}
\overrightarrow{\theta}_{pol}(s)=\overrightarrow{\theta}_{pol}(0)+\int_{0}%
^{s}\vec{\kappa}_{pol}(s)ds\label{tethapol}%
\end{equation}
Now we can compute the polymorphic variance from the correlation function
$G_{pol,x}(s^{\prime},s)=\left\langle (\theta_{pol,x}(s^{\prime})-\theta
_{x}(0)(\theta_{pol,x}(s)-\theta_{x}(0))\right\rangle .$ Note that by symmetry
$G_{pol,y}(s^{\prime},s)=G_{pol,x}(s^{\prime},s).$ From Eqs. \ref{tethapol}%
,\ref{CC} we readily obtain for the full $4$-block model%
\begin{align}
G_{pol,x}(s^{\prime},s) &  =\frac{\kappa_{1}^{2}P^{2}}{2}\int_{0}^{s^{\prime}%
}\int_{0}^{s}ds_{1}ds_{2}\cos\left(  q_{0}(s_{2}-s_{1}\right)  )\nonumber\\
&  \exp\left(  -\frac{\left\vert s_{2}-s_{1}\right\vert }{\widetilde{\xi}%
b}\right)  .\label{Gpol}%
\end{align}
where for a free microtubule (not angularly constrained at the ends) we have
to integrate over the arbitrary initial angle $\varphi\in\left[
0,2\pi\right]  .$ The variance can now be computed and we obtain :
\begin{align}
V_{pol}(s^{\prime}-s) &  =\frac{2\kappa_{1}^{2}P^{2}\xi}{1+\xi^{2}q_{0}^{2}%
}\left\{  \left\vert s^{\prime}-s\right\vert -\frac{1-\xi^{2}q_{0}^{2}}%
{1+\xi^{2}q_{0}^{2}}\xi\right.  \nonumber\\
&  +\frac{\xi e^{-\left\vert s^{\prime}-s\right\vert /\xi}}{1+\xi^{2}q_{0}%
^{2}}\left(  \left(  1-q_{0}^{2}\xi^{2}\right)  \cos\left(  q_{0}(s^{\prime
}-s)\right)  \right.  \nonumber\\
&  \left.  \left.  -2q_{0}\xi\sin\left(  q_{0}\left\vert s^{\prime
}-s\right\vert \right)  \right)  \right\}  \label{VARI}%
\end{align}
with $\xi=\widetilde{\xi}b.$ From Eq. \ref{VARI} the persistence length
$l_{pol}$ and thus $l_{p}$ can be deduced. The polymorphic variance Eq.
\ref{VARI} has the following generic asymptotic behaviors : At large distance
$s>>\xi,$ the variance $V_{pol}(s)\approx2\kappa_{1}^{2}P^{2}\frac{\xi}%
{1+\xi^{2}q_{0}^{2}}s$ scales linearly with $s$ indicating that
$\overrightarrow{\theta}_{pol}$ is performing a simple (angular) random walk.
At such a scale the microtubule looses its "coherent nature" and is replaced
by a collection of uncorrelated segments. So not surprisingly we recover the
classical results of a semiflexible chain again. In this asymptotic regime the
effective persistence length reaches saturation with a renormalized constant
value $l_{p}\left(  \infty\right)  =1/\left(  l_{pol}^{-1}+l_{B}^{-1}\right)
$\ where
\begin{equation}
l_{pol}\left(  \infty\right)  =\frac{1+\xi^{2}q_{0}^{2}}{\xi\kappa_{1}%
^{2}P^{2}}\label{lpinfinit}%
\end{equation}
At short distance such that $s<<$min($\xi,q_{0}^{-1}$) the variance has a
quadratic behavior $V_{pol}(s)\approx\kappa_{1}^{2}P^{2}s^{2}$. In this
regime, the polymorphic fluctuations are completely dominated by purely
"classical" semiflexible chain fluctuations and $l_{p}\left(  0\right)
=l_{B}.$ Starting from this value of $l_{p}$ at $s=0$, polymorphic
fluctuations begin to contribute reducing the persistence length $l_{p}\left(
s\right)  \approx(\frac{1}{l_{B}}+\frac{\kappa_{1}^{2}P^{2}s}{2})^{-1}$. This
behavior is universal and independent of $\xi$ and $q_{0}^{-1}.$

In the intermediate regime with $s$ of the order of $\xi$, the behavior of
$V_{pol}(s)$ depends on the value of $\xi q_{0}$ - a kind of "helix coherence"
parameter. When $\xi q_{0}\leq1$ (low helix coherence) oscillations in Eq.
\ref{VARI} are damped and the persistence length $l_{p}\left(  s\right)  $ is
monotonously deceasing from $l_{B}$ to the constant $l_{p}\left(
\infty\right)  $ with $l_{pol}\left(  \infty\right)  \approx(\xi\kappa_{1}%
^{2}P^{2})^{-1}$ obtained from Eq. \ref{lpinfinit} (cf. Fig.4). This
corresponds also to the situation of no internal twist $q_{0}=0.$

In the most interesting (high helical coherence) regime $\xi q_{0}>1,$ the
behavior of $V_{pol}(s)$ is a combination of two effects: an oscillation with
wave length $\lambda=2\pi q_{0}^{-1}$ originating from the helical nature of
the microtubule, but which is now damped by the presence of thermally induced
defects (due to a finite $\xi$) reducing the coherence of the helix and
enhancing the linearly growing behavior (the random walk). As a consequence
for high helical coherence $l_{p}\left(  s\right)  $ displays three different
regimes (cf Fig.4):

I. In the limit of very short , $l_{p}\left(  s\right)  \approx(\frac{1}%
{l_{B}}+\frac{\kappa_{1}^{2}P^{2}s}{2})^{-1}$ that attains a global minimum at
$s_{\min}\approx\pi q_{0}^{-1}$

II. For intermediate length values $s_{\min}<s\lesssim$ $\xi$, the total
persistence length displays a non-monotonic oscillatory behavior of period
$\lambda$ with damped amplitude around a nearly linearly growing average
reflecting the polymorphic fluctuation of the helix.

III. For distances $s>>\xi,$ the oscillations in Eq. \ref{VARI} are completely
damped as the helix forgets its "coherent nature" on these scales and reaches
the limiting value $l_{p}^{\ast}\left(  \infty\right)  =1/\left(  l_{pol}%
^{-1}+l_{B}^{-1}\right)  $\ with $l_{pol}\left(  \infty\right)  $ given by Eq.
\ref{lpinfinit}.

\begin{figure}[ptbh]
\begin{center}
\includegraphics[
width=3.1081in ]{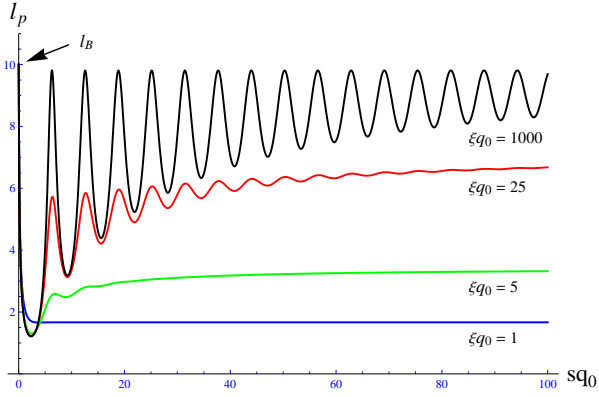}
\end{center}
\caption{Representative plots of the angular persistence length $l_{p}$ versus
distance along contour $s$ for intrinsic elastic bending stiffnes $l_{B}=10\mu
m$ (chosen arbitrary and small for pictorial convenience) and different values
of scaled correlation length $\xi q_{0}.$}%
\end{figure}

\section{High cooperativity limit and uniform states}

In the previous paragraphs we considered the thermodynamic limit
$L=Nb\rightarrow\infty$ for which the correlation length $\xi$ is always
smaller that the total length $L$ \cite{Footnote1}. It also interesting to
consider the opposite regime $\xi>>L$. In this case of large cooperativity we
may take formally the limit $J\rightarrow\infty$ . Consequently the two states
variables $\sigma^{i}$ become uniform all along their respective block axes
$i=1..4$. The decoupling of the full MT lattice into 2 independent $2\times2$
sub-blocks implies the equality of the energy and entropy of the two
independent 2-sub-blocks sets so that we can focus on only a single
sub-lattice made of two blocks, say $1$ and $2.$ The corresponding polymorphic
energy is $E=\frac{B\kappa_{1}^{2}}{2}L\left[  (\frac{\pi}{2}\gamma-1)\left(
\sigma^{1}+\sigma^{2}\right)  +2\sigma^{1}\sigma^{2}\right]  $ and its
partition function reads now $Z=1+2\exp\left(  -2(\widetilde{\gamma
}-\widetilde{B})L/b\right)  +\exp\left(  -4\widetilde{\gamma}L/b\right)  .$
For two blocks there are then only 4 possible states with different
probabilities of realization $\rho(\sigma^{1},\sigma^{2})=\exp(-\beta E)/Z$
that are: $\rho(0,0)=1/Z$ for the straight unstrained state, $\rho
(1,1)=\exp\left(  -4\widetilde{\gamma}L/b\right)  /Z$ for the straight
prestrained state and finally $\rho(0,1)=\rho(1,0)=\exp\left(  -2(\widetilde
{\gamma}-\widetilde{B})L/b\right)  /Z$ for the two polymorphic helical states.
The average value of the spin and the correlation functions are given
respectively by $\left\langle \sigma^{k}\right\rangle =\rho(1,0)+\rho(1,1)$
for $k=1..4$ and $\left\langle \sigma^{1}\sigma^{2}\right\rangle =\left\langle
\sigma^{3}\sigma^{4}\right\rangle =\rho(1,1)$.

Going back to the full lattice model made of four blocks the polymorphic order
parameter is then given $P^{2}(L)=2(\left\langle \sigma^{1}\right\rangle
-\left\langle \sigma^{1}\sigma^{2}\right\rangle +\left\langle \sigma
^{3}\right\rangle -\left\langle \sigma^{3}\sigma^{4}\right\rangle
)=4\rho(1,0)$. The angular variance at position $s$ and $s^{\prime}$ can be
easily computed and we find $V(s^{\prime}-s)=\frac{\kappa_{1}^{2}P^{2}%
(L)}{q_{0}^{2}}\left(  1-\cos\left(  q_{0}(s^{\prime}-s\right)  )\right)  $
which obviously corresponds to the limit of infinite $\xi$ of Eq. \ref{VARI}
but with the additional factor $P^{2}(L)$ responsible for the length
dependence of the variance. The weight of the different possible
configurations depends on $\widetilde{\gamma}$ and the discrimination between
them is more pronounced with growing length. This can be easily understood
from the polymorphic entropy $\Sigma_{\sigma}$ which remarkably is non
extensive. Note that $\Sigma_{\sigma}$ is always twice the entropy of the
individual 2-block-sub-lattices. For the asymptotic limiting case of short
$L\rightarrow0$ all $4\times4$ configurations of the full lattice have the
same probability and therefore $\Sigma_{\sigma}(0)=2k_{B}\ln4.$ For larger
$L,$ the entropy will continuously change to reach a constant value depending
on the dominant configurations that are selected by $\gamma.$ For values of
$\left\vert \widetilde{\gamma}\right\vert >\widetilde{B}$, we see on Fig. 5
that $P(L)$ goes to zero with growing values of $L$. In this regime there is
only one configuration (the same for the two sublattices) $\rho(0,0)$ or
$\rho(1,1)$ depending on the sign of $\widetilde{\gamma}$ and $\Sigma_{\sigma
}(L)\rightarrow0.$ Interestingly for $\widetilde{\gamma}=\pm\widetilde{B},$
there is an additional configuration for a given sub-lattice with the same
probability $\rho(0,1)=\rho(1,1)=\rho(0,0)=1/Z,$ so that for the 4-blocks we
have $3\times3$ configurations and $\Sigma_{\sigma}(L)\rightarrow2k_{B}\ln3.$
This coexistence leads to a state which is the superposition between straight
and curved states with different probability such that $P^{2}(L)=\frac
{4}{3+\exp\left(  -4\widetilde{B}L/b\right)  }\approx\frac{4}{3}$ that is
(quasi) length independent. When $\left\vert \widetilde{\gamma}\right\vert $
is very close to $\widetilde{B}$ , the evolution of $\Sigma_{\sigma}(L)$
versus $L,$ can be slow and therefore there is a regime of lengths where the
the entropy is non-extensive and where the straight states have comparable
free energy with the helical one. In this regime the microtubule will
fluctuate between its almost degenerate straight and (two) curved states.

For $\left\vert \widetilde{\gamma}\right\vert <\widetilde{B},$ there are for a
given sub-lattice two dominant degenerate configurations $(0,1)$ and $(1,0),$
so that the number of configuration for the full lattice model is $2\times2$
and $\Sigma_{\sigma}(L)\rightarrow2k_{B}\ln2$ with growing $L.$ In this regime
the mean curvature is built up progressively with the length $L,$ to reach
-faster for smaller value of $\left\vert \widetilde{\gamma}\right\vert $ - its
maximum value $P=\sqrt{2}.$ Here also when $\left\vert \widetilde{\gamma
}\right\vert $ is very close to $\widetilde{B}$ the realization of the curved
state can be slow with growing $L.$ \begin{figure}[ptbh]
\begin{center}
\includegraphics[
width=3.1081in ]{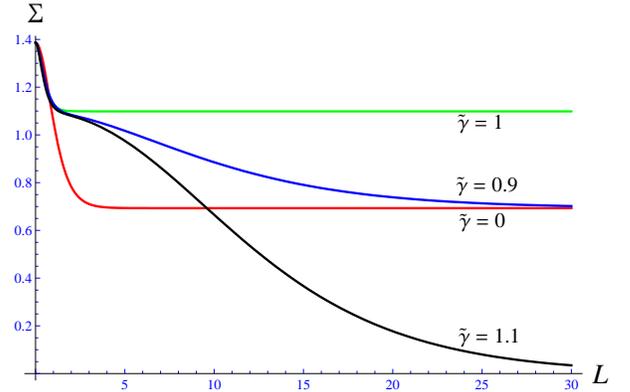}
\end{center}
\caption{Polymorphic entropies $\Sigma$ (four-block model) versus the
microtubule's length $L$ for different values of conformational bias
$\widetilde{\gamma}$ and stiffness $\widetilde{B}=1.$ }%
\end{figure}\begin{figure}[ptbh]
\begin{center}
\includegraphics[
width=3.1081in ]{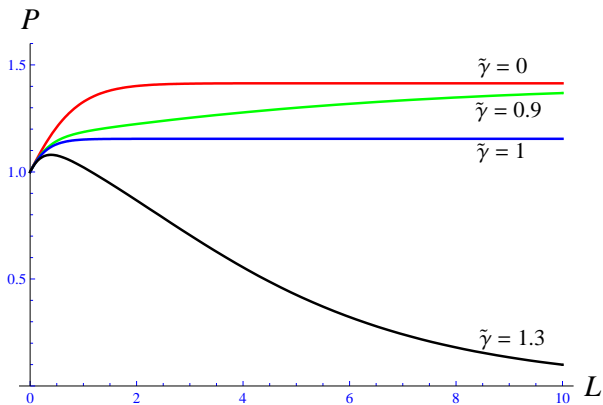}
\end{center}
\caption{Plots of the polymorphic polarization $P$ (four-block model) versus
the microtubule's length $L$ for different values of $\widetilde{\gamma}$ and
with $\widetilde{B}=1.$ We observe that microtubule curvature can be build up
progressively with growing MTs. }%
\end{figure}

\section{Angular power spectrum}

Other information on the equilibrium properties of the microtubule can be
deduced from the analysis of its Fourier mode distribution. To do so we
express the microtubule shape $\overrightarrow{\theta}(s)$ as a superposition
of Fourier modes $\overrightarrow{\theta}(s)=\overrightarrow{\theta}%
(0)+\sqrt{2/L}\sum_{n=1}^{+\infty}\overrightarrow{a}_{n}\cos(q_{n}%
s)+\sqrt{2/L}\sum_{n=1}^{+\infty}\overrightarrow{b}_{n}\sin(q_{n}s)+\left(
\theta_{x}(L)-\theta_{x}(0)\right)  s/L$ with the wave vector $q_{n}%
=\frac{2n\pi}{L}$ and where we add the "ramp term" $\left(  \theta
_{x}(L)-\theta_{x}(0)\right)  s/L,$ as $\overrightarrow{\theta}(s)$ is not a
periodic function in general, i.e., $\overrightarrow{\theta}(L)\neq
\overrightarrow{\theta}(0)$ \cite{Footnote2}. Decomposing the shape as
$\overrightarrow{\theta}(s)=\overrightarrow{\theta}_{pol}(s)+\overrightarrow
{\theta}_{el}(s)$ we readily see that the equilibrium square of the amplitude
of the elastic Fourier modes $G_{el}(q_{n})\equiv\left\langle \overrightarrow
{a}_{el,n}^{2}\right\rangle =\left\langle \overrightarrow{b}_{el,n}%
^{2}\right\rangle =\frac{2}{l_{B}q_{n}^{2}}$ has the usual thermally driven
elastic bending spectrum of a wormlike chain. On the other hand from Eq.
\ref{Gpol} the polymorphic mode distribution $G_{pol}(q)$ in the limit $L$
large (with $q_{n}\rightarrow q$ continuous) can be deduced :%

\begin{equation}
G_{pol}(q)=\frac{2\kappa_{1}^{2}P^{2}\xi}{q^{2}}\cdot\frac{1+\xi^{2}%
(q^{2}+q_{0}^{2})}{1+\xi^{4}(q^{2}-q_{0}^{2})^{2}+2\xi^{2}(q^{2}+q_{0}^{2}%
)}\label{Gp}%
\end{equation}
As shown in Fig. 6, when the correlation length $\xi$ is of order of the helix
pitch $\lambda=2\pi/q_{0}$, the function $G_{pol}(q)$ departs very much from
the wormlike chain bending spectrum $G_{el}(q)$ and displays a non-monotonous
shape with a peak at $q_{0}$. This peak is more pronounced and sharper for
larger value of $\xi q_{0}$ (helix more coherent) and disappears when $\xi
q_{0}\leq1$. This is the main Fourier mode signature of the -polymorphic
fluctuating- helicoidal nature of the microtubule. For long wavelength modes
$q<<q_{0},$ the spectrum has a semiflexible like behavior $G_{pol}%
(p)\approx\frac{2}{l_{pol}\left(  \infty\right)  q^{2}}$ reflecting the
incoherence of the helix at this scale, with the persistence length given by
Eq. \ref{lpinfinit}. For short wavelength modes $q>>q_{0},$ the behavior
becomes quarticaly decreasing $G_{pol}(p)\approx\frac{2\kappa_{1}^{2}}{\xi
q^{4}}$ and polymorphic fluctuations are strongly damped at short distance in
agreement with high cooperativity at this scale. \begin{figure}[ptbh]
\begin{center}
\includegraphics[
width=3.1081in ]{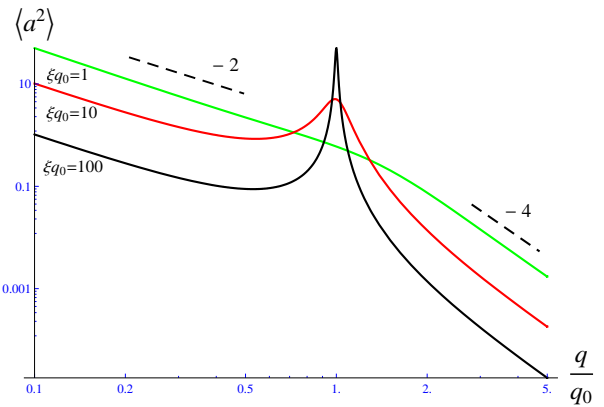}
\end{center}
\caption{Log-Log plot of the polymorphic Fourier mode distribution
$G_{pol}(q)$ for different values of $\xi q_{0}.$ Besides a characteristic
peak at $q_{0}$ one observes a cross-over between a small $q$ regime with
scaling $q^{-2}$ and a large $q$ regime with a $q^{-4}$ scaling. }%
\end{figure}

Quantitative comparison with experiments at this point seems difficult,
because Fourier analysis of microtubules in-vitro was mainly performed on
microtubules strongly confined into an almost 2-D geometry. This experimental
necessity (originating from optical microscopy sample flatness requirements)
is expected to induce artifacts as in this case, within our helical model, the
microtubules would likely not fully equilibrate between their different
equivalent curved states as large barriers will appear under confinement. In
agreement with this argument the experimental Fourier modes of microtubules
systematically reveal some type of "frozen-in" curvature which is much larger
than the fluctuations around it \cite{GITTES} - in particular for longer
wavelength modes. This is in so far striking as no clear argument for large
systematic built-in curvature in microtubules (other than the here described)
is evident, indicating that this curvature is in fact the very character of
the microtubule elasticity. Similarly in vivo, it was found that growing
microtubules have large scale frozen-in curvature \cite{Brangwyne}. This
phenomenon was interpreted in terms of a random interaction of the growing
microtubule tip with the surrounding cytoskeleton. Instead it could be that
the curvature has an intrinsic origin (like in the present model) but is not
allowed to fully equilibrate due to constraints present in the cytoplasm.

\section{Conclusion and Outlook}

We have investigated the thermodynamics of a toy model for cooperatively
switching (polymorphic) biofilaments, whose paradigm example is believed to be
the microtubule. The polymorphism of the microtubule's subunits and the short
range cooperative interaction leads to properties that are very different from
usual biofilaments. In particular the ground state itself is\ not a single
fixed shape but is found to consist of a set of degenerate helicoidal
configurations. At finite temperature, the fluctuations of the subunits create
defects that smoothen the perfect helix and can even destroy it on very long
scale where the filament retrieves a typical worm-like chain behavior. Another
peculiar characteristic of the polymorphic tube is the length dependence of
the persistence length. The Fourier spectrum turns out also to be markedly
different from classical biofilaments. While it has a typical semiflexible
chain behavior at long scale the Fourier modes amplitude are enhanced for
modes around the characteristic scale of the helix. At very short scales the
modes are strongly dampened due to the strong cooperative dimer interaction.
We believe that the Fourier spectrum will be of some value to discriminate
between different models in the analysis of future experiments (as in
\cite{Pampaloni}) where confinement is absent and microtubules can thus
rearrange and equilibrate.

The model developed here was inspired and tailored to microtubules. However it
can be adapted to other polymorphic filaments with similar cross-sectional
symmetry (circular rod or tube). Polymorphic sheets, rods and tubes are not
uncommon in nature with the best documented example the bacterial flagellum
\cite{Flagellum}. In the latter case the polymorphic states are likely very
rigid (or coherent in our terminology) on thermal energy scales with
correlation lengths $\xi$ larger than filament's length $L$. Rearrangements
between equivalent lattice states could then face high barriers in the
flagellum and suppress the here described polymorphic fluctuations on
measurement time scales. However other filaments might be even better
candidates for soft polymorphic tubes (or rods) and deserve a closer
investigation. Observations of unusual flaring-up of peculiar helical modes in
bacterial pili (cf. Fig. 1 in \cite{Pili}) as well as in actin filaments
cooperatively interacting with drugs like cofilin (cf. Fig. 1c in
\cite{ActinCofilin}) visually suggest the presence of soft polymorphic mode dynamics.


\begin{thebibliography}{99}
\bibitem {Footnote1}The periodic boundary conditions on the spin inherent to
the tranfer matrix method implies that Eq. \ref{CC} is valid for $L>>\xi.$ For
$L$ of order $\xi,$ we need to consider free boundary conditions for which we
have computed the spin correlations numerically for growing values of $N.$ A
difference between numerics and the analytical - asymptotic $N\rightarrow
\infty$ expression Eq. \ref{CC} comes from the correlation function that
decreases suddently slightly faster than $\exp(-\left\vert i-j\right\vert
/\widetilde{\xi})$ for $i$ or $j$ near the ends. This boundary effect is
rapidly negligeable and already for $N=100$ a very good correspondance between
the numerics and \ref{CC} was found.
\bibitem {Footnote2}We have tried the more conventional decomposition
$\overrightarrow{\theta}(s)=a_{0}+\sqrt{2/L}\sum_{n=1}^{+\infty}%
\overrightarrow{a}_{n}\cos(\frac{n\pi}{L}s)$ as in \cite{GITTES} and, from it,
computed the variance $V_{pol}(s)$ and found that is does not give Eq.
\ref{VARI}. Fourier decomposition around a non-trivial ground state is a
complex issue with mathematical "underwater mines" that we will discuss elsewhere.
\end{thebibliography}
\end{document}